\documentclass{aa}  

\usepackage{graphicx}
\usepackage{epstopdf}
\usepackage{txfonts}
\usepackage{hyperref}
\usepackage{xcolor}
\begin{document}

\title{Close  galaxy pairs with accurate photometric redshifts}

   \author{Facundo Rodriguez
\thanks{facundo.rodriguez@unc.edu.ar}\inst{1,2} \and Elizabeth Johana Gonzalez \inst{1,2,3} \and Ana Laura O’Mill \inst{1,2} \and Enrique Gaztañaga \inst{4,5} \and Pablo Fosalba \inst{4,5},  Diego Garc\'ia Lambas \inst{1,2}
\and Mar Mezcua\inst{4,5} \and Małgorzata Siudek \inst{6,7}
}

\institute{Universidad Nacional de Córdoba. Observatorio Astronómico de Córdoba. Córdoba, Argentina
         \and
           CONICET. Instituto de Astronomía Teórica y Experimental. Laprida 854, X5000BGR, C\'ordoba, Argentina
         \and
          Centro Brasileiro de Pesquisas F\'isicas, Rio de Janeiro, RJ 22290-180, Brasil
         \and
           Institute of Space Sciences (ICE, CSIC), 08193 Barcelona, Spain
         \and  
          Institut d'Estudis Espacials de Catalunya (IEEC), 08034 Barcelona, Spain 
         \and
         Institut de F\'{\i}sica d'Altes Energies (IFAE), The Barcelona Institute of Science and Technology, 08193 Bellaterra (Barcelona), Spain
         \and
         National Centre for Nuclear Research, ul. Hoza 69, 00-681 Warszawa, Poland
          }


 
  \abstract
   {Studies of galaxy pairs can provide valuable information to jointly understand the formation and evolution of galaxies and galaxy groups. Consequently, taking into account the new high precision photo-z surveys, it is important to have reliable and tested methods that allow us to properly identify these systems and estimate their total masses and other properties. }
   {In view of the forthcoming Physics of the Accelerating Universe Survey (PAUS) we propose and evaluate the performance of an identification algorithm of projected close
   isolated galaxy pairs. We expect that the photometric selected systems can adequately reproduce the observational properties and the inferred lensing mass - luminosity relation of a pair of truly bound galaxies that are hosted by the same dark matter halo.  }
   {We develop an identification algorithm that considers the projected distance between the galaxies, the projected velocity difference and an isolation criteria in order to restrict the sample to isolated systems. We apply our identification algorithm using a mock galaxy catalog that mimics the features of PAUS. To evaluate the feasibility of our pair finder, we compare  the identified photometric samples with a test sample that considers that both members are included in the same halo. Also, taking advantage of the lensing properties provided by the mock catalog, we apply a weak lensing analysis to determine the mass of the selected systems. }
   {Photometric selected samples tend to show high purity values, but tend to misidentify truly bounded pairs as the photometric redshift errors increase. Nevertheless, overall properties such as the luminosity and mass distributions are successfully reproduced. We also accurately reproduce the lensing mass - luminosity relation as expected for galaxy pairs located in the same halo. }
   {}


   \maketitle
%

\section{Introduction}

In a hierarchical formation scenario, pairs of galaxies can provide the first stages 
of the formation of  massive systems. Close galaxy pairs, in particular, 
can be useful to study galaxy evolution since interactions between the pair members 
are common and can leave to significant changes in their physical properties 
\citep{Toomre1972,Patton2016,Hernandez2005,Woods07, Ellison2010,Mesa2014}. Thus, these systems can be
considered as major merger progenitors given the expected incorporation of the stellar populations 
of the satellite galaxies into the most massive galaxy \citep[e.g., ][]{Patton2000, Lin2004}. 
Other works have also considered central-satellite systems \citep[e.g., ][]{Norberg2008}.
Recently, \cite{Ferreras2019} found that satellites with similar stellar velocity dispersions have older stellar population when orbiting around massive primaries, supporting the galaxy bias scenario in this regime. In spite that galaxy 
pairs are important to study the halo assembly scenario \citep{Gao2005}
as well as galaxy morphology transformations, there are only few studies on the subject.  
Having a reliable sample of galaxy pairs where both 
galaxies belong to the same halo is a challenge that can provide important clues on 
the formation of larger structures and galaxy evolution. 

Observational galaxy pair catalogs are mainly constructed considering
a limiting velocity difference, $\Delta V$, computed according to spectroscopic redshift information, 
and a limiting projected distance between the member galaxies, $r_p$ 
\citep{Lambas2003,Lambas2012,Chamaraux2016,Ferreras2017,Nottale2018}. 
Therefore, they are mainly based on spectroscopic galaxy surveys and are limited to 
relatively small physical scales since these surveys typically cover relatively small areas, 
but with a high galaxy density \citep[e.g.,][]{Davis2003, Lilly2007}.
On the other hand, the identification of these systems based only
on photometric information can be difficult
given the uncertainty of redshift estimates. \citet{LopezSanjuan2015} 
identify close pairs using photometric redshifts based
on the ALHAMBRA survey, by considering the full probability distribution functions of the sources 
in redshift space. They select the pairs setting $r_p =100$ kpc and $\Delta V = 500$ km s$^{-1}$. 
Using this approach they can successfully reproduce merger fractions and
rates in agreement with those derived from spectroscopic surveys. This result shows that 
these particular systems can be identified using photometric data
in order to recover physical properties comparable with those obtained
using on spectroscopic samples.

The masses of dark matter halos contain valuable information regarding the evolution of the 
systems and are key to understanding the connection
between the luminous and the dark matter content.
In particular, mass determinations of halos hosting galaxy pairs can contribute significantly to a better understanding of the joint evolution of galaxies and groups.  
Also, halo masses for these systems can be used in the context of the Halo Occupation Distribution (HOD) models to follow galaxy formation and system evolution.
Virial masses of galaxy pairs have usually been determined according to the dynamics, using different methods
\citep{Nottale2018,Chengalur1996,Peterson1979,Faber1979}. 
These methods are affected by projection effects
of the parameters of the virial mass estimation, $\Delta V$ and $r_p$. Moreover, it should be stressed that this approach only gives information
about the total mass enclosed by the member galaxies. 

On the other hand,
weak gravitational lensing has proved to be an efficient technique to derive total halo masses of galaxy systems \citep[e.g. ][]{Wegner2011,Dietrich2012,Jauzac2012,Umetsu2014,Jullo2014,Gonzalez2018}. The main
shortcoming of this approach is that 
the detection of weak-lensing signals is difficult given that the small shape distortions of background galaxies are substantially limited by their intrinsic ellipticity dispersions \citep{Niemi2015}.
Taking into account that isolated galaxy pairs
are likely to be low mass systems, a weak lensing signal is expected for individual 
pairs resulting in a low signal-to-noise mass estimate. 
However, by analyzing galaxy pairs using stacking techniques it is possible to derive accurate mean mass estimates. 
These techniques have been implemented to study low-mass galaxy 
groups and to obtain average properties of the combined systems  \citep[e.g. ][]{Leauthaud10,Melchior13,Rykoff08,Foex14, Chalela2017, Chalela2018,Pereira2018}.
Recently, stacking techniques have been successfully applied in order to 
derive average masses of galaxy pairs, finding general agreement with HOD predictions and  other works that link mass to luminosity \citep{Gonzalez2019}. These authors obtain higher lensing masses for pairs with signatures of interaction, red members and high luminosity. 
They note, however, that these results can also be affected by the inclusion of interlopers alone the line-of-sight for blue, non-interacting members which could bias low the mass estimates. Therefore, testing the identification algorithms
is important in order to interpret the results properly. 

Here we develop and test an algorithm for the identification
of nearly equal mass close galaxy pairs using simulated data in order to predict
observable results for the Physics of the Accelerating Universe Survey
\citep[PAUS,][]{Padilla2019,Eriksen2019} 
and  the Canada France Hawaii Telescope Lensing Survey \citep[CFHT, ][]{Heymans2012,Miller2013}. 
The purpose of the identification algorithm
is to obtain photometrically selected systems that reproduce the lensing
masses that would be derived for isolated galaxy pairs that reside in the same dark matter halo. We focus our work on the upcoming 
PAUS data to select the systems, and the CFHT lensing catalog 
to derive the system masses. PAUS aims to observe $\sim100$\,deg$^2$ down to $i_{AB} < 22.5$,
reaching a volume of 0.3 (Gpc/h)$^3$ with several million redshifts \citep{Padilla2019}. 
The PAUS camera takes images of the sky with 40 narrow bands that cover the wavelength range from 
4500 \AA \ to 8500 \AA \ at 100 \AA \  intervals. These images are combined with 
existing deep broad band photometry to obtain high precision photo-z \citep{Eriksen2019}. 

Our work is organized as follows: In Section \ref{MICE} 
we describe MICE simulation on which we base our identification algorithm,
as well as  testing and  lensing analysis.
In Section \ref{Iden}, we introduce the criteria for the galaxy pair selection algorithm, and  describe the different resulting samples. 
In Section \ref{Lens} we describe the lensing analysis implemented, 
first for a general sample of halos to test our lensing techniques, 
and then applied to the pair samples. Finally, in section \ref{Sum} we present the summary and our conclusions.
\section{MICE simulation}
\label{MICE}

For our analysis we use version 2 of the Marenostrum Institut de Ciències de l'Espai (MICE) simulation\footnote{\href{http://maia.ice.cat/mice/}{http://maia.ice.cat/mice/}} \citep{Fosalba2015,Carretero2015,Fosalba2015b,Crocce2015}. This is a cosmological N-body dark matter only simulation containing $4096^3$ dark-matter particles of mass $m_p = 2.93\times 10^{10}h^{-1}M\odot$ in a box-volume of $3072^3 (Mpc/h)^3$ run using the \textsc{GADGET-2} code.  Halos are resolved down to a few $10^{11}M\odot h^{-1}$ using a hybrid Halo Occupation Distribution (HOD) and Halo Abundance Matching (HAM) technique for galaxy modeling resulting in a total number of approximately $5 \times 10^8$ galaxies. The simulation also included a sky footprint of $90 \times 90\,$ deg$ ^2$ filling an octant of sky $(f_{sky}= 0.125)$ up to redshift $z = 1.4$ as well as several galaxy properties \citep[following][]{Carretero2015}. The assumed cosmology is a flat concordance $\Lambda CDM$ model with $\Omega _m = 0.25$, $\Omega _{\Lambda} = 0.75$, $\Omega _b = 0.044$, $n_s = 0.95$, $\sigma _8 = 0.8$ and $h = 0.7$ consistent with WMAP 5-year data. 

This simulation has the advantage of having lensing parameters, such as shear and convergence, as well as magnified magnitudes and angular position, computed for each synthetic galaxy \citep{Fosalba2015b}. These parameters are calculated following the `onion Universe' approach described in \citet[see ][]{Fosalba2008}, that is equivalent to ray-tracing techniques in the Born approximation. Lensing values are assigned to each galaxy according to its 3D position and do not include shape-noise.

Data acquisition is performed using the CosmoHub platform\footnote{\href{https://cosmohub.pic.es/home}{https://cosmohub.pic.es/home}} \citep{Carretero2017}. The selected fields used for the analysis include: the unique halo and galaxy ID,
\texttt{unique\_gal\_id} and \texttt{unique\_halo\_id}; the sky position of the galaxies, \texttt{ra\_gal}
and \texttt{dec\_gal}; the shear parameters, \texttt{gamma1} and \texttt{gamma2}; the observed galaxy redshift
\texttt{z\_cgal}; the flag that identifies the galaxy as central or satellite, \texttt{flag\_central};
the logarithmic halo mass, \texttt{lmhalo} and the magnitudes corrected for evolution.

We constrain our analysis to a region of four patches of 5$\times$5 deg$ ^2$ in order to have a comparable sky coverage to the upcoming PAUS data. 

\section{Galaxy pair identification}
\label{Iden}
In this section, we present the algorithm adopted to identify galaxy pair candidates from a mock catalogue with  photometric data. The algorithm searches for pairs by selecting galaxies within a given projected distance ($r_p$) and a radial velocity difference ($\Delta V$). We optimize the procedure using bright galaxies as centers taking into account the magnitude difference between the galaxy member candidates. 

We propose a simple method for galaxy pair identification, similar to those applied to spectroscopic surveys, considering the  uncertainties of photometric redshift. We have tested the method using expected uncertainties in high precision photometric redshift surveys as well be discuss in more detail in section \ref{Photo pairs}.

\subsection{The algorithm}

The proposed identification algorithm follows the traditional approach to search for galaxy pairs using $r_p$ and $\Delta V$ parameters. Also, we improve the identification by taking into account the photometric characteristics of the system: requiring that the pair has a galaxy brighter than a certain magnitude limit and establishing a limiting magnitude difference between the members ($\Delta m$). Finally, we consider an isolation criterion to ensure that the pairs are not part of a larger system.

First, we select all the galaxies brighter than an absolute SDSS $r-$band magnitude $-19.5$ as potential pair centers. Then, we search for another galaxy fainter than the center, within $r_p = 50$kpc and a given $\Delta V$ difference that depends on the photometric redshift error. 
The identified systems also have to satisfy an apparent magnitude difference of $\Delta m <2$. This last criterion, together with the adopted luminosity threshold of the centers, guarantees the identification of real pairs that are neither a faint satellite, nor orphan system. Also, limiting the apparent magnitude difference ensures that the identified members are nearly equal-mass galaxies which are expected to merge \citep{Kitzbichler2008,Jian2012,Moreno2013}, constituting close major-merger pairs. Finally, an isolation criteria is applied so that there is no other galaxy within $5r_p$.

Since the photometric redshift errors mainly affect the determination of the pair velocity difference, we consider three samples and analyze the most appropriate $\Delta V$ values for each case. In the following subsections, we will discuss this approach further.

\subsection{Samples}

 To assess the efficiency of our galaxy pair identification algorithm, we select pairs in the patches that belong to the same halo and meet the criteria defined above. Then, we test the reliability to recover them using photometric data. For this aim, we use three samples with different photometric redshift accuracies.
 Galaxy pair identification considering the observational characteristics of PAUS data and as our main goal is to determine mass profiles using weak lensing analysis, we restrict our identification to the redshift range $0.2<z<0.6$, taking into account the redshift distribution of PAUS.

 \subsubsection{True pairs}

\begin{figure}[h]
    \includegraphics[width=0.51\textwidth]{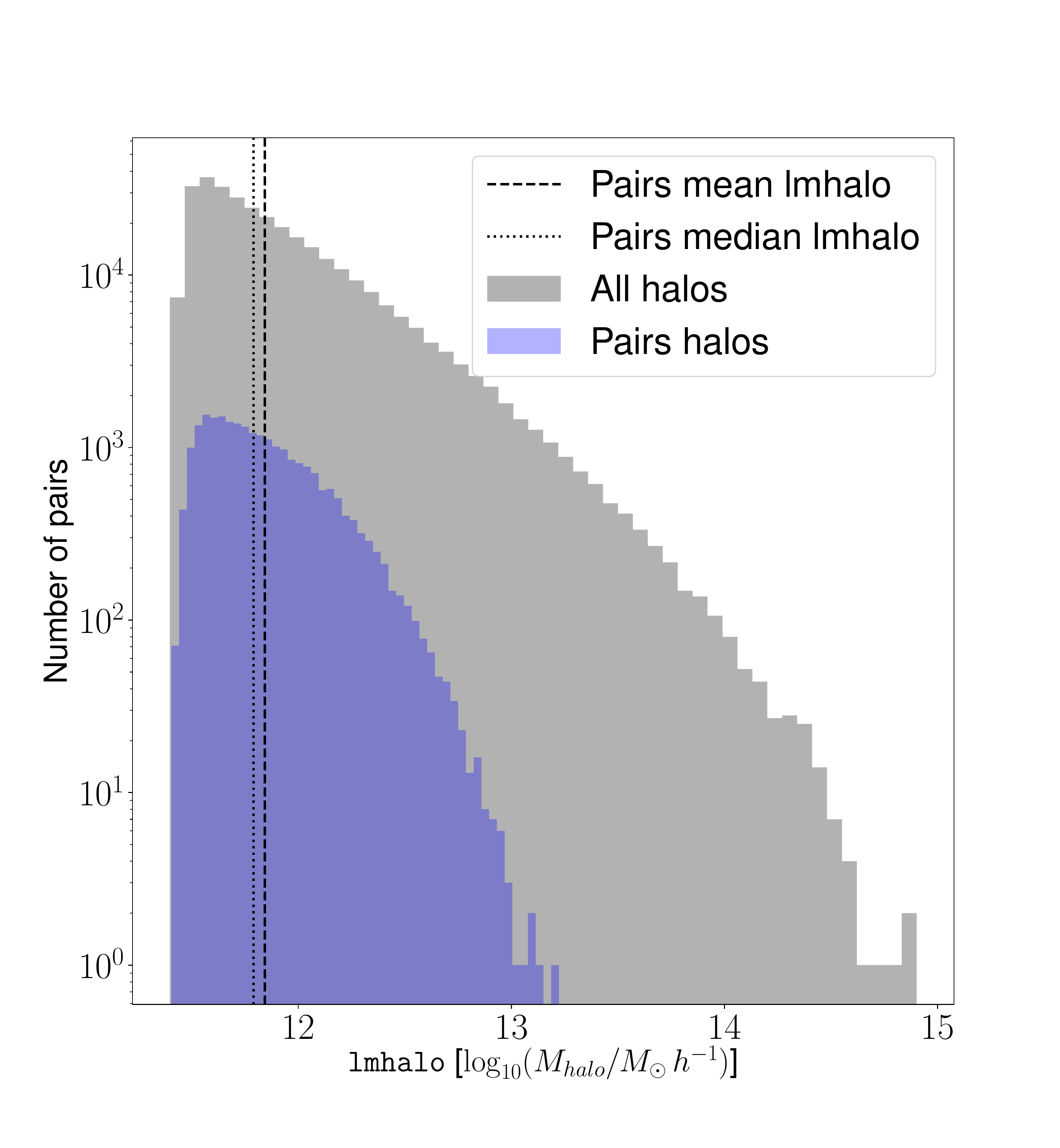}
    \caption{Halo mass distribution for all
    halos (grey) and halos hosting \textit{true pairs} (blue). Dashed and dotted lines correspond the mean ($10 ^{11.84}$ M$_{halo}/$M$_\odot\,h^{-1}$) and median ($10 ^{11.79}$ M$_{halo}/$M$_\odot\,h^{-1}$) values, respectively.}
    \label{fig:massdist}
\end{figure}

We define a control sample of galaxy pairs that satisfies the selection criteria and that also belongs to the same halo, which hereafter we call the \textit{true pairs} sample. This is accomplished by requiring $r_p<50$kpc, $\Delta V < 350$ km s$^{-1}$, one member galaxy with an absolute $r-$band magnitude brighter than $-19.5$ (central), a relative magnitude difference $\Delta m <2$ and an isolation criteria $5r_p$,
plus the restriction that both galaxies reside in the same dark matter halo.
We obtain  24\,523 \textit{true pairs} within the four 5$\times$5 deg$ ^2$ regions. Thus, despite the fact that our identification algorithm does not explicitly require that one of the galaxy pair members is a halo central galaxy (\texttt{flag\_central} $= 0$), all the identified pairs have a central as one of the member galaxies. This is expected taking into account that the pairs reside in low mass halos and one of the members has to be a luminous galaxy. Figure \ref{fig:massdist} shows the mass distribution of the host halos of the \textit{true pairs}. The mass range is as expected for this type of isolated system and we do not observe significant differences between the mass distributions of the different angular regions. 

\begin{figure*}[h]
    \centering
    \includegraphics[scale=0.3]{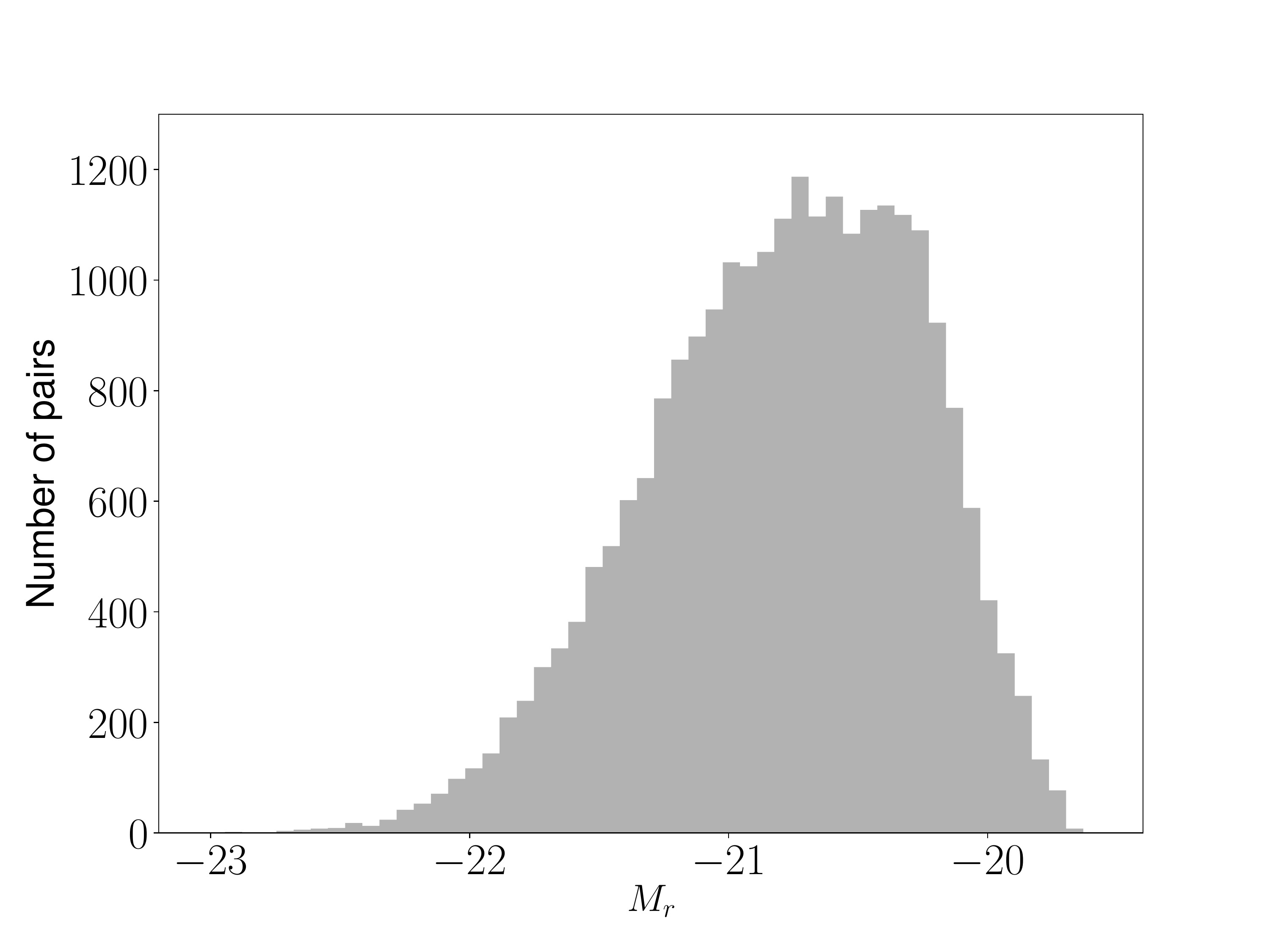}   \includegraphics[scale=0.3]{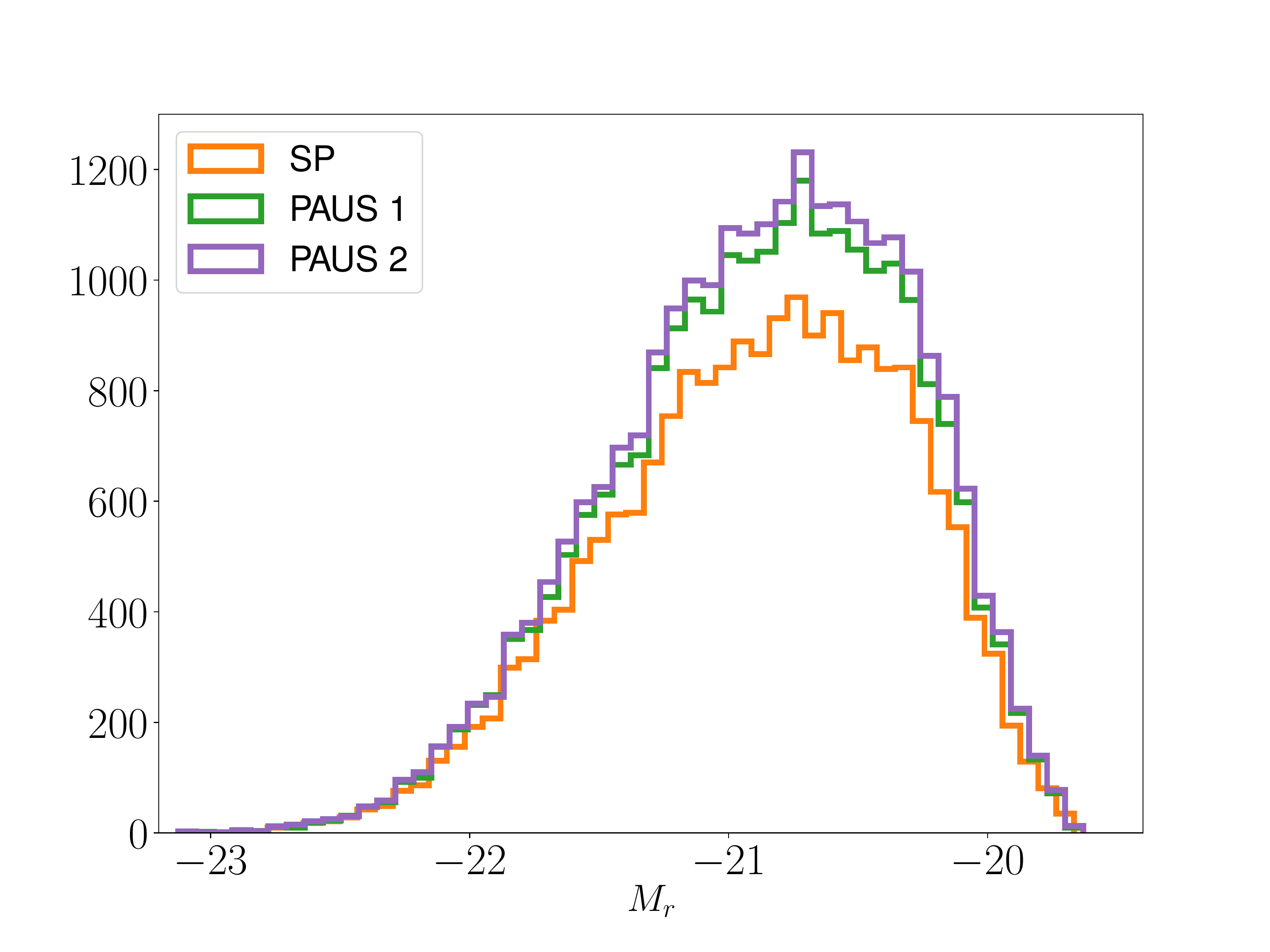}
    \caption{Distribution of the total luminosity for \textit{True pairs} sample (\textit{left}) and for the photometric selected samples (\textit{right}). SP is the sample with an standard photometric uncertainty ($\delta_z = 0.01$) while PAUS\,1  and PAUS\,2 are  the two samples with high precision photometric redshift ($\delta_z = 0.002$ and $\delta_z = 0.0037$, respectively). }
    \label{fig:lumdist2}
\end{figure*}

In Figure \ref{fig:lumdist2} (left panel) present the total absolute $r-$band magnitude distribution for the \textit{true pairs} sample, $M_r = -2.5\log_{10}(L1+L2)$, where $L1$ (central) and $L2$ (companion) are the $r-$band luminosity of the pair members that will be compared later to the photometrically identify samples.

\begin{figure}
    \centering
    \includegraphics[scale=0.65]{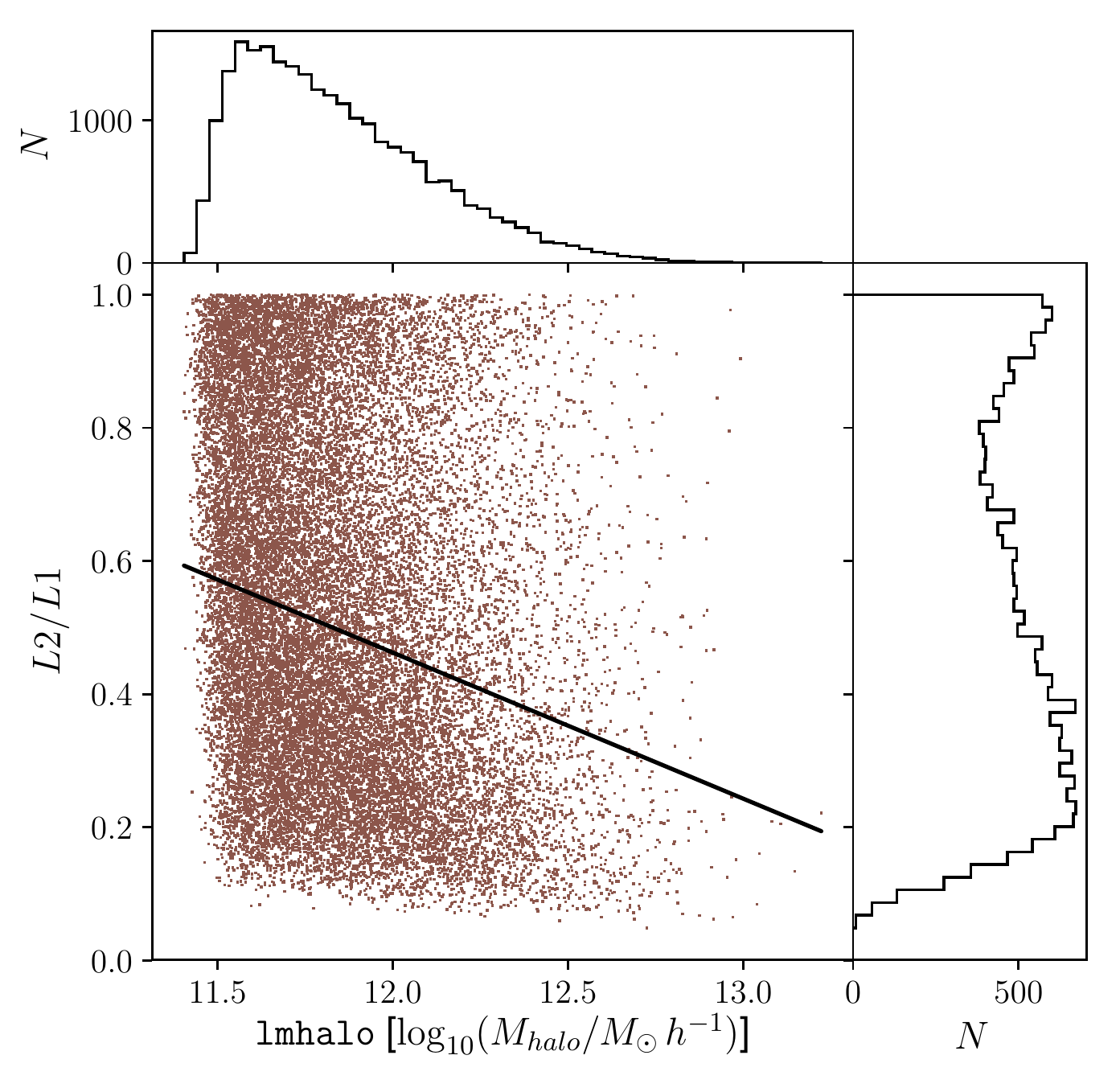}
    \caption{Luminosity ratio of the \textit{True pairs} members versus the mass of the halo where the pairs reside. The black line
    shows the fit of the median values of the $L2 / L1$ obtained in 10 percentiles of halo mass. }
    \label{fig:ratiomass}
\end{figure}
\begin{figure}[h]
    \centering
    \includegraphics[scale = 0.65]{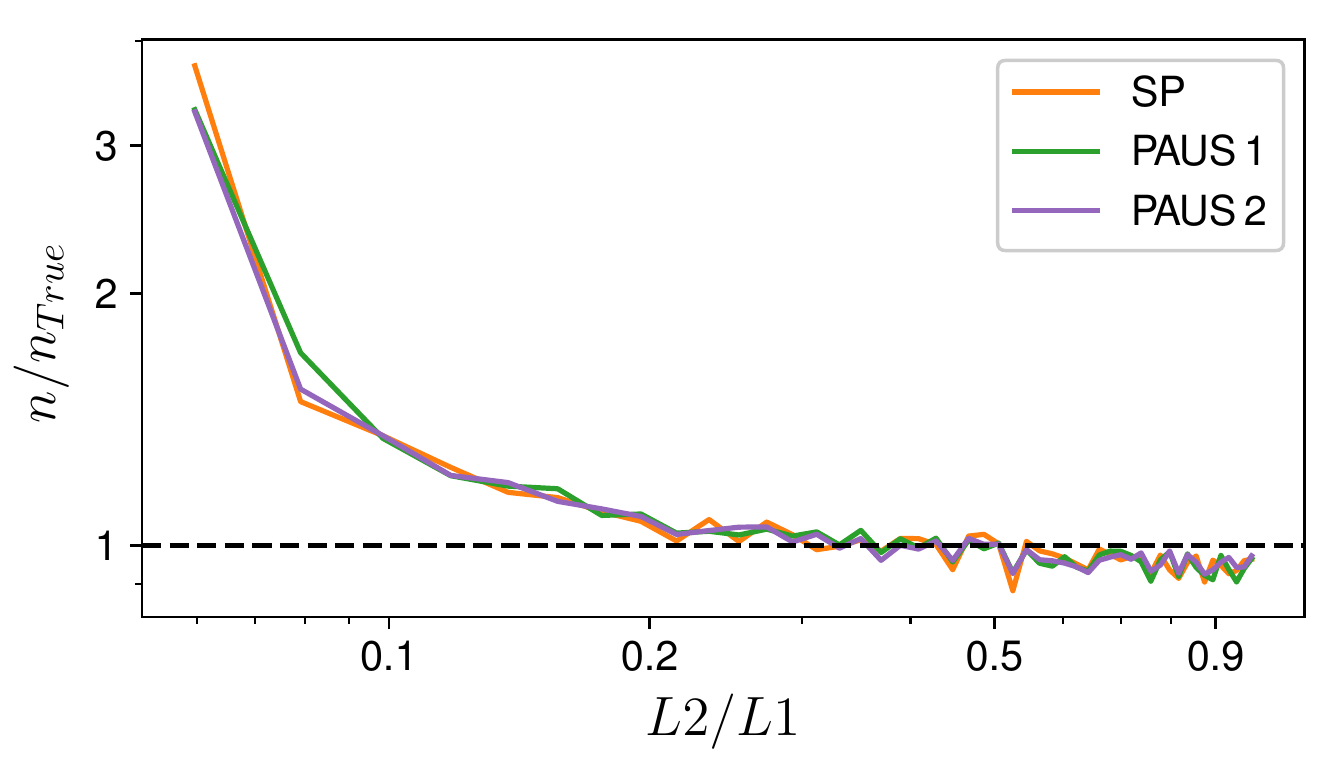}
    \caption{Ratio between the fraction of pairs selected within $L2/L1$ bins for each photometric sample, $n$, and for the  \textit{true pairs}, $n_{True}$. $L2/L1$ distributions are in excellent agreement with the derived for the \textit{true pairs} when considering $L2/L1 > 0.2$. It is important to highlight that pairs with $L2/L1 < 0.2$ constitute $\sim 10 \%$ for all the selected samples.}
    \label{fig:coclumdist}
\end{figure}

The left panel of Fig. \ref{fig:ratiomass} shows the distribution of the \textit{true pairs} luminosity ratio, $L2/L1$, where a bi-modality can be seen with a maximum at 0.25, i.e. pairs that have a central galaxy four times brighter than their companion. 
When comparing $L2/L1$ ratios with halo mass of the pairs, we find that pairs with members of similar luminosity tend to reside in halos with larger masses .

 \subsubsection{Photometric pairs}
 \label{Photo pairs}
Once the \textit{true pairs} sample is obtained, we apply our algorithm in the selected patches of the MICE simulation considering the galaxies with an imposed redshift error that reproduces the photometric data. To mimic the observational catalogs, we add to the  \texttt{z\_cgal} parameter suitable photometric errors . We define two samples with high precision photometric redshift errors that follow the values expected in PAUS, and a third sample, SP, with lower precision, standard photometric uncertainties ($\delta_z = 0.01$). 

Following  \cite{Eriksen2019}, we consider the uncertainties of two samples with high precision photometric redshift as $\delta_z \times (1 +z)$, with $\delta_z = 0.002$ and $\delta_z = 0.0037$, labelled as PAUS\,2 (a better quality sample) and PAUS\,1 (which corresponds to the typical expected photo-z precision for PAUS), respectively. For each sample we take into account a galaxy photometric redshift, $z_{phot}$, taken from a Gaussian distribution with \texttt{z\_cgal} as the mean and the expected uncertainties as the $1-\sigma$ standard deviation. 

We identify pairs in the three samples with the different redshifts uncertainties taking into account a compromise between purity and completeness for setting the $\Delta V$ parameter. In this procedure we simply evaluate that the velocity difference of both galaxies is less than the given $\Delta V$ value taking into account the assigned photometric redshift $z_{phot}$  (i.e. c|$z_{phot,1}- z_{phot,2}$|$<\Delta V$).

Purity, $P$, quantifies the chance of pair members to reside in the same halo:
\begin{equation}
    P=N_{i}/N_{True},
\end{equation}
where $N_{i}$ is the number of identified pairs in which both members reside in the same halo and $N_{True}$, is the total number of \textit{true pairs}. Thus, high values of $P$ exclude a significant number of spurious pairs in the photometric selected sample.

On the other hand, the halo completeness, $C$, quantifies if the halos where \textit{true pairs} reside are identified as pairs:
\begin{equation}
    C=N_{i}/N_{iden},
\end{equation}
where $N_{iden}$ is the total number of identified pairs. $C$ provides information about the total pairs that we can recovered with our procedure.

To set the $\Delta V$ threshold for each photometric sample we test several values in order to maximize the number of identified pairs with the highest $C$ and $P$ parameters. The general properties of the obtained \textit{photometric pair} samples are listed in Table\,\ref{tab:samples}. It can be noticed that, as the photometric redshift error increases, completeness is more affected than purity, therefore larger redshift errors tend to lose \textit{true pairs} at a higher rate than to identify galaxies that reside in different halos. 

\begin{table}
\caption{General properties of the identified \textit{Photometric pairs} samples. }
    \label{tab:samples}
    \centering
    \begin{tabular}{c c c c c c}
        \hline
        \hline
         Sample & $\delta_z$ & $\Delta V [km s^{-1}]$ & Number & P & C \\ 
               &             &                        & of pairs & & \\
         \hline
        SP         & 0.01   & 8\,500 & 20\,508 & 0.82 & 0.68 \\
        PAUS\,1  & 0.0037 & 3\,500 & 24\,061 & 0.85 & 0.82 \\
        PAUS\,2  & 0.002  & 2\,500 & 25\,135 & 0.88 & 0.86 \\
        \hline
\end{tabular}
\medskip
\begin{flushleft}
\end{flushleft}    
\end{table}

It is important to note that photometric samples show very similar observational properties as the \textit{true pairs}, both in total luminosity distribution as in members luminosity ratio (Fig. \ref{fig:lumdist2} and \ref{fig:coclumdist}). Nevertheless, photometric samples tend to include more pairs with higher luminosity compared with the \textit{true pairs} and with lower $L2/L1$. 

\section{Lensing Analysis}
\label{Lens}

In order to predict the lensing signal 
associated with the different galaxy pair samples \footnote{Note that in this work
we refer as a galaxy pair as the lens system instead of a lens-source pair, 
commonly used in other analysis.},
we use the lensing properties provided by MICE. 
We first assess and validate the mass determination  for a sample of pure halos 
binned according to the FOF halo mass. Then,
we apply the same analysis to the three photometric redshift galaxy pair samples: SP, PAUS\,1 and PAUS\,1. 

We first describe the stacking technique to derive total masses.
We select source galaxies,
that is galaxies affected by lensing, 
taking into account the available 
\textit{shear} catalogs in MICE.
Then we present the results obtained for 
the total halo samples and for the galaxy pair samples. 

\subsection{Stacking techniques}

Gravitational lensing distorts 
the shape of background galaxies that lie behind galaxy systems. The induced shape distortion is 
quantified by the shear parameter, $\gamma$,  that can be related to
the projected density distribution of the galaxy system. 
Shear estimates are obtained in observations according to the measured 
ellipticities of the galaxies. Nevertheless, since galaxies are not 
intrinsically round, the observed ellipticity is a combination of
the intrinsic galaxy ellipticity and the lensing shear effect. 
The dispersion of intrinsic ellipticities introduces noise in  shear estimates, known as `shape noise',
that is proportional to the inverse square root of the number of source galaxies.

Stacking techniques are commonly used to derive the total
mass of the composite lenses considered \citep[e.g. ][]{Leauthaud10,Melchior13,Rykoff08,Foex14, Chalela2017, Chalela2018,Pereira2018,Gonzalez2019}.
The implementation of this methodology allows us to
increase the number of source galaxies,
which decreases the shape noise and results in better estimates of the total mass. Moreover, the resulting projected density
distribution is softened, reducing the impact
of the substructures present in the halos.

Application of the stacking methodology consists of the combination 
of many lenses by averaging the measured distortions of source galaxies.
In the case of spherical symmetry, the average 
of the tangential shear component, $\tilde{\gamma}_{T}(r)$, 
in an annulus of physical radius $r$ is related to the projected density contrast, 
$\Delta\tilde{\Sigma}$, defined as:
\begin{equation}
    \tilde{\gamma}_{T}(r) \times \Sigma_{\rm crit} = \bar{\Sigma}(<r) - \bar{\Sigma}(r),
\end{equation}
where $\bar{\Sigma}(<r) $ and $\bar{\Sigma}(r)$ are the average projected 
mass distribution within a disk, and in a ring of radius $r$, 
respectively. $\Sigma_{\rm crit}$ is the critical density defined as:
\begin{equation}
    \Sigma_{\rm{crit}} = \dfrac{c^{2}}{4 \pi G} \dfrac{D_{OS}}{D_{OL} D_{LS}},
\end{equation}
that considers the geometrical configuration of the observer-lens-source system, 
through the angular diameter distances
between the observer to the source, $D_{OS}$, the observer to the
lens $D_{OL}$ and the lens to the source $D_{LS}$, respectively. On the other hand, the average of the 
shear component tilted by $\pi/4$, called the cross component, $\tilde{\gamma}_{\times}(r)$
should be zero and is used to test for systematic effects. 

We can combine the lensing signal for a number of lenses, 
$N_{Lenses}$, and derive the projected 
density contrast profile, by averaging the tangential component of the shear: 
\begin{equation}
\langle \Delta \tilde{\Sigma}(r) \rangle = \frac{\sum_{j=1}^{N_{Lenses}} \sum_{i=1}^{N_{Sources,j}} \gamma_{T,ij} \times \Sigma_{\rm{crit},ij}}{N_{total\,sources}},
\end{equation}
where $N_{Sources,j}$ and $N_{total\,sources}$ are 
the total number of sources located at a distance $r \pm \delta r$
for the $j$th lens and for the whole sample of lenses considered, respectively. 
$\Sigma_{\rm{crit},ij}$ is the critical density for the $i$th source of the $j$th lens. 

Density contrast profiles are obtained by considering logarithmic equispaced radial bins,
from $r_{\mathrm{in}} = 350$\,kpc, taking into account the lensing resolution of MICE v2.0 (pixel\_size = 0.43 arcmin), up to  $r_{\mathrm{out}}$. The value of $r_{\mathrm{out}}$ is computed according to the average halo mass of the  lenses
in order to avoid the region where the two-halo term becomes significant. For this, we use the relation presented by \citet{Simet2017} between the richness
and the upper-limit radius combined with their mass-richness relation, taking
into account the halo mass provided by MICE. In the case of the galaxy pairs samples we
estimate this radius according to this relation and fix its value to $r_{\mathrm{out}} = 1.0$\,Mpc.

\subsection{Source galaxy selection}

We select MICE source galaxies taking into account the characteristics of the CFHTLenS survey, which
provides weak lensing catalogs in regions that overlap with the PAUS data (in fact CFHTLenS is the reference catalog for PAUS forced aperture narrow band photometry that is used to measure accurate photometric redshifts).
This survey is based on deep multicolor data and spans 154 square degrees distributed in four patches 
W1, W2, W3 and W4 (63.8, 22.6, 44.2 and 23.3 deg$^2$ respectively). 
Lensing catalogs include photometric redshift estimates, $Z_B$, computed by \citet{Hildebrandt2012}
using the Bayesian photometric redshift software \textsc{bpz} \citep{Benitez2000,Coe2006} which is
used for the source galaxy selection.
We also apply 
a cut in the \textsc{ODDS} parameter, a measure of the quality of the redshift estimate. This parameter varies between 0 and 1, where galaxy samples with larger \textsc{ODDS} values have a lower fraction of outliers \citep[See e.g. ][]{Eriksen2019}.

We compute the surface
density of background galaxies that would be expected using these lensing catalogs
by selecting the galaxies from the CFHTLenS catalog
with $0.2 < Z_B < 1.3$ and with \textsc{ODDS}$> 0.5$.
With these requirements, the density of sources is $\sim 7 $\,galaxies\,arcmin$^{-2}$
considering the masking regions. Taking these issues into account, we first
select from MICE catalog the galaxies with $i^\prime_{AB} < 24.7$,
which is the CFHTLenS limiting magnitude \citep{Heymans2012}. Then
source galaxies are randomly selected in order to obtain the same density 
as that expected from CFHTLens data at the same
redshift range. Each halo at redshift $z_H$ is considered as a lens 
 and we select the sources as those galaxies with \texttt{z\_cgal} $> z_H + 0.1$. This last criterion
is usually applied in a stacking lensing analysis \citep{Leauthaud2017,Pereira2018,Chalela2018,Gonzalez2019}.

In our analysis, we consider two source galaxy samples,
one \textit{noisy}  and the other \textit{noise-free}.
For the \textit{noisy} sample, we simulate the observational noise
by adding to the shear a Gaussian random value with zero average ellipticity
and dispersion $\sigma_e = 0.28$. This value corresponds to the 
measured ellipticity dispersion of the CFHTLenS survey, and includes both 
intrinsic ellipticity dispersion and measurement noise \citep{Simon2015}.
On the other hand, the \textit{noise-free} sample considers the original shear parameters provided by MICE.

Errors in the density contrast profiles based on the \textit{noise-free} source sample
are obtained for each radial bin according to the
standard error, considering the dispersion of the individual profiles obtained
for all the $N_{Lenses}$ halos included in the stacking. In the case of the profiles
derived using the \textit{noisy} source sample, we estimate the error as
$\sigma_e/\sqrt{N}$, where $N$ is the total number of source galaxies considered in the
radial bin.

\begin{figure*}
\includegraphics[scale=0.53]{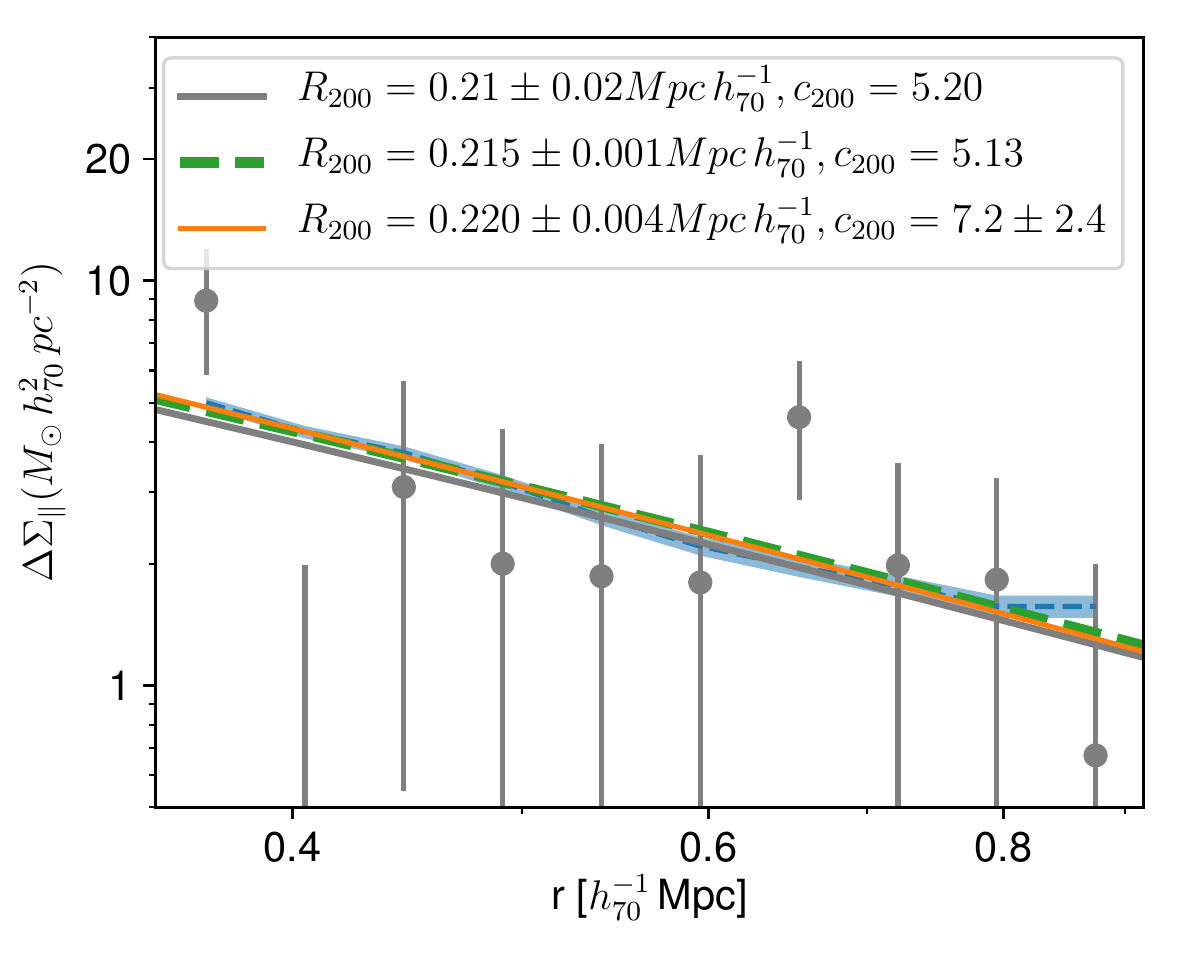}
\includegraphics[scale=0.53]{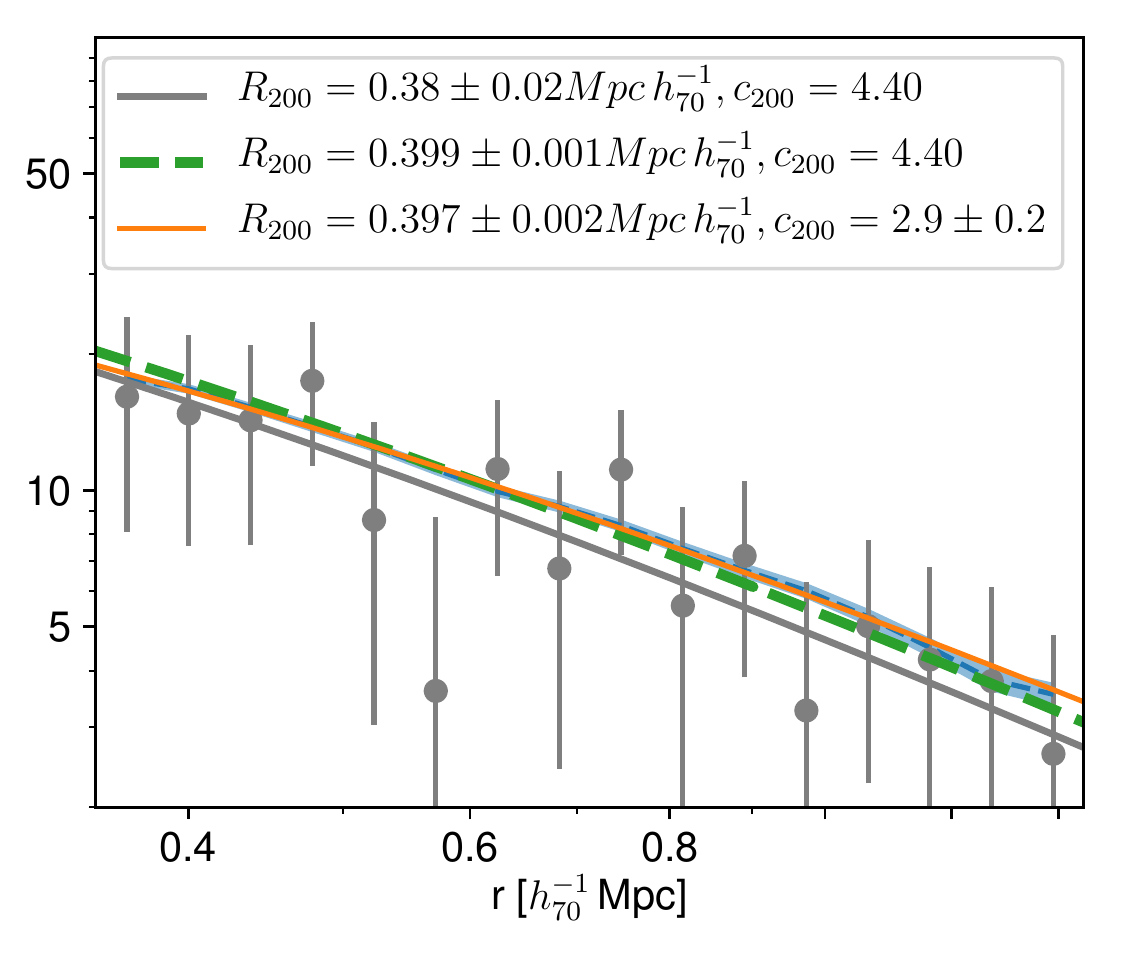}
\includegraphics[scale=0.53]{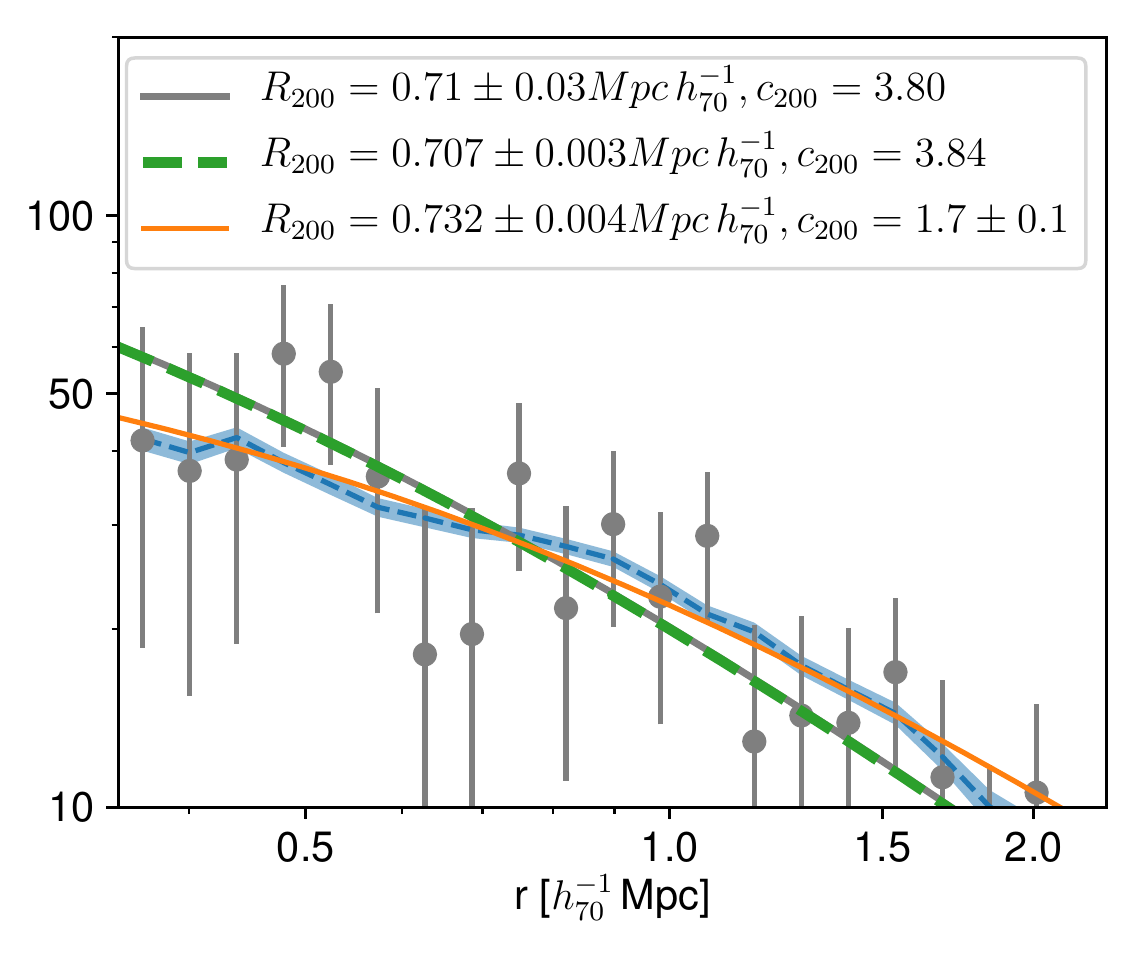}
\caption{Average density contrast profiles for three halo mass bins, from left to right
\texttt{lmhalo} $\in$ [12.10,12.30), [12.90,13.10), [13.70,13.90)
log(M/M$_\odot$). Density contrast derived from the \textit{noise-free} source
sample is shown in blue, and the best-fit NFW fitting the $c_{200}$ (solid orange line)
and using \citet{Duffy2008} relation (dashed green line). Grey points
corresponds to profiles obtained according to the \textit{noisy}
source sample and its corresponding best-fit NFW in solid grey line. 
$h_{70}$ corresponds to $h=0.7$.}
\label{fig:profiles}
\end{figure*}

\subsection{Modelling the lensing signal}

Halo masses are obtained by fitting the computed contrast density profiles
using an NFW density distribution model \citep{Navarro1997}. 
This profile depends on two parameters, $R_{200}$, which is the radius 
that encloses a mean density equal to 200 times the critical density 
of the universe, and a dimensionless concentration parameter,
$c_{200}$. This density profile is given by:
\begin{equation} \label{eq:nfw}
\rho(r) =  \dfrac{\rho_{\rm crit} \,\delta_{c}}{(r/r_{s})(1+r/r_{s})^{2}}, 
\end{equation}
where $\rho_{\rm crit}$ is the critical density of the universe at the average redshift
of the lenses, $r_{s}$ is the scale radius, $r_{s} = R_{200}/c_{200}$ and $\delta_{c}$
is the cha\-rac\-te\-ris\-tic overdensity of the halo:
\begin{equation}
\delta_{c} = \frac{200}{3} \dfrac{c_{200}^{3}}{\ln(1+c_{200})-c_{200}/(1+c_{200})}.  
\end{equation}
The mass within $R_{200}$ can be obtained as 
\mbox{$M_{200}=200\,\rho_{\rm crit} (4/3) \pi\,R_{200}^{3}$}. 
Lensing formulae for the NFW density profile were taken from \citet{Wright2000}. 

There is a well known degeneracy between $R_{200}$ and $c_{200}$ that can
be broken only if we consider information about the density distribution in the inner radius. For the 
profiles based on the \textit{noise-free} source sample we fit both parameters.
We also compute the masses by using a fixed mass-concentration relation 
$c_{200}(M_{200},z)$, derived from simulations by \citet{Duffy2008}: 
\begin{equation}
c_{200}=5.71\left(M_{200}/2 \times 10^{12} h^{-1}M_\odot\right)^{-0.084}(1+z)^{-0.47},
\end{equation}
where we take $z$ as the mean redshift value of the lens sample. In the case of the
\textit{noisy} sample, we only fit $R_{200}$ and consider the 
previous \citet{Duffy2008} relation, since $R_{200}$ and $c_{200}$  can not be simultaneously 
constrained given the observed profile uncertainty.

\subsection{Lensing results for halos}

We apply the described lensing analysis considering as lenses
all the halos satisfying the same redshift range as the identified
galaxy pairs, $0.2 < $\texttt{z\_cgal}$ < 0.6$,
and with an \texttt{lmhalo} $> 11.5$\,log(M/M$_\odot$). With these criteria
the total sample of lenses includes 231\,970 halos. We split the sample 
according to the \texttt{lmhalo} parameter on 15 evenly spaced bins of 0.2 dex width.  
Derived profiles for three halo mass bins are shown in 
Fig.\,\ref{fig:profiles}. 

\begin{figure}
    \centering
    \includegraphics[scale=0.65]{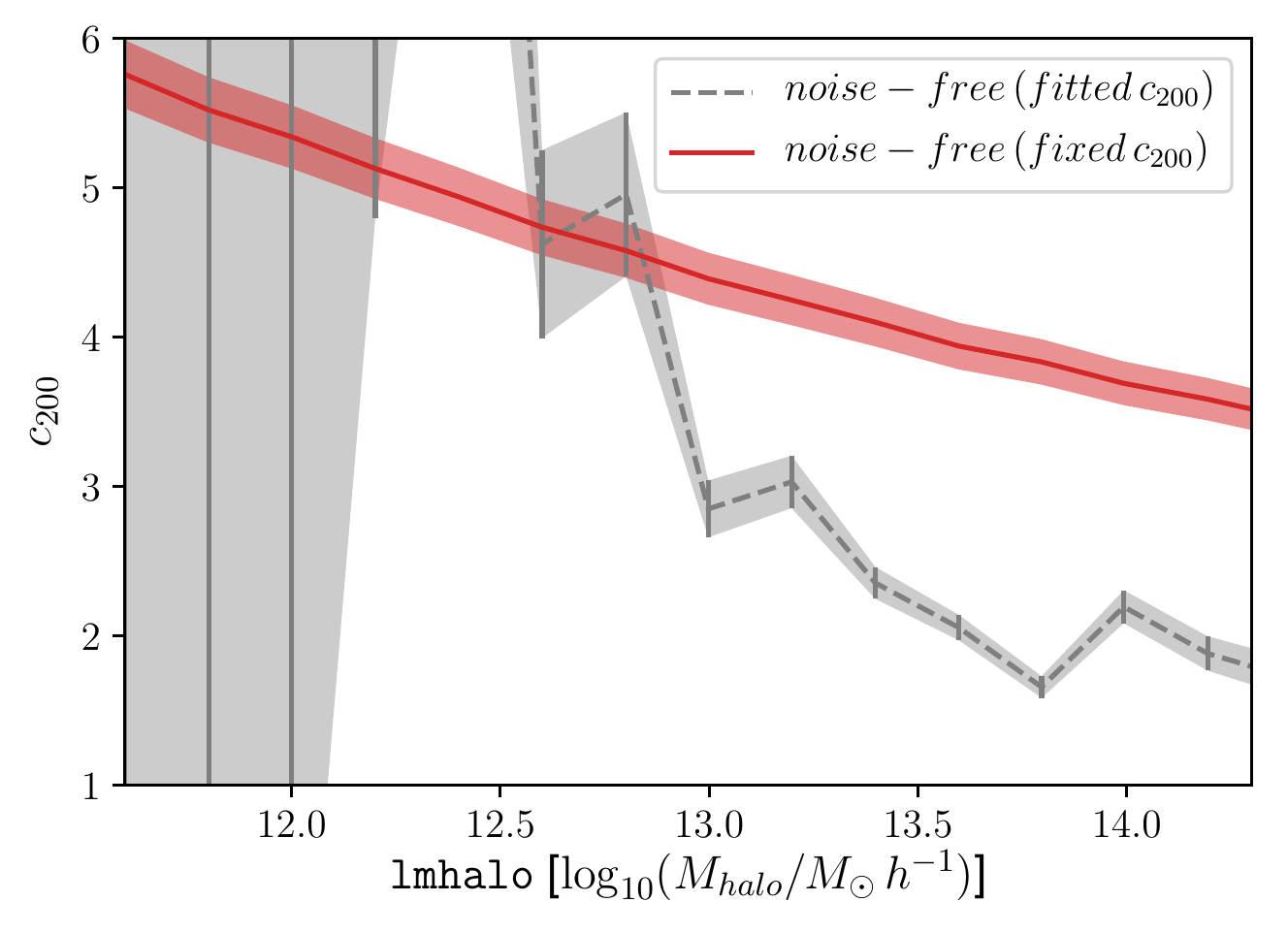}
    \caption{Derived concentration parameters from the profiles based on
    the \textit{noise-free} source sample (dashed grey line) versus halo masses, 
    compared with the concentrations obtained using the \citet{Duffy2008} relation
    (solid red line). The grey area corresponds to the errors in the fitted
    concentrations. The red area is computed according to the errors in the 
    parameters of \citet{Duffy2008} relation. Fitted concentrations are well constrained for halos with 
    masses \texttt{lmhalo} $> 12.5$\,log(M/M$_\odot$), however predicted concentrations using \citet{Duffy2008} relation are $\sim 2$ times the values obtained by fitting this parameter.}
    \label{fig:concentration}
\end{figure}

We evaluate the derived halo concentrations by comparing the fitted
$c_{200}$ parameter for the profiles obtained by using the \textit{noise-free} source sample. 
In Fig. \ref{fig:concentration} we show the fitted concentration parameters
together with those predicted according to the \citet{Duffy2008} relation,
as a function of halo mass. Concentration
values cannot be accurately determined for \texttt{lmhalo} $< 12.5$\,log(M/M$_\odot$). 
Poorly constrained concentrations can be due to the lack of information 
in the inner regions of the density profiles. This is more important for 
low mass halos since changes in the profile slope corresponding to different concentrations
are significant at smaller radii.
For \texttt{lmhalo} $> 12.5$\,log(M/M$_\odot$),
fitted concentrations tend to be lower than the predicted values. 
This can be seen from the highest mass
bin shown in Fig.\,\ref{fig:profiles}, where the derived profile from 
the  \textit{noise-free} source sample  flattens at small radius
compared to the best-fit NFW model using the \citet{Duffy2008} relation. This result
is in agreement with that obtained by \citet{Lopez2014}. By analyzing the stacked 3D density profiles of halos in MICE,  
he derives best-fitting concentrations that are lower than predicted from other literature relations. 
In spite of the similarity of the cosmological parameters of the MICE and \citet{Duffy2008} simulations, 
for \texttt{lmhalo} $\geqq 13$\,log(M/M$_\odot$)
the fitted concentrations are roughly half of those predicted 
using the \citet{Duffy2008} relation. 
These observed differences could be due to a higher softening length of the MICE simulation
($l_{soft} =50$\,h$^{-1}$\,kpc , which is 100 times larger than 
the softening used by \citet{Duffy2008}) which can lead to less concentrated halos. 

\begin{figure}
    \centering
    \includegraphics[scale=0.65]{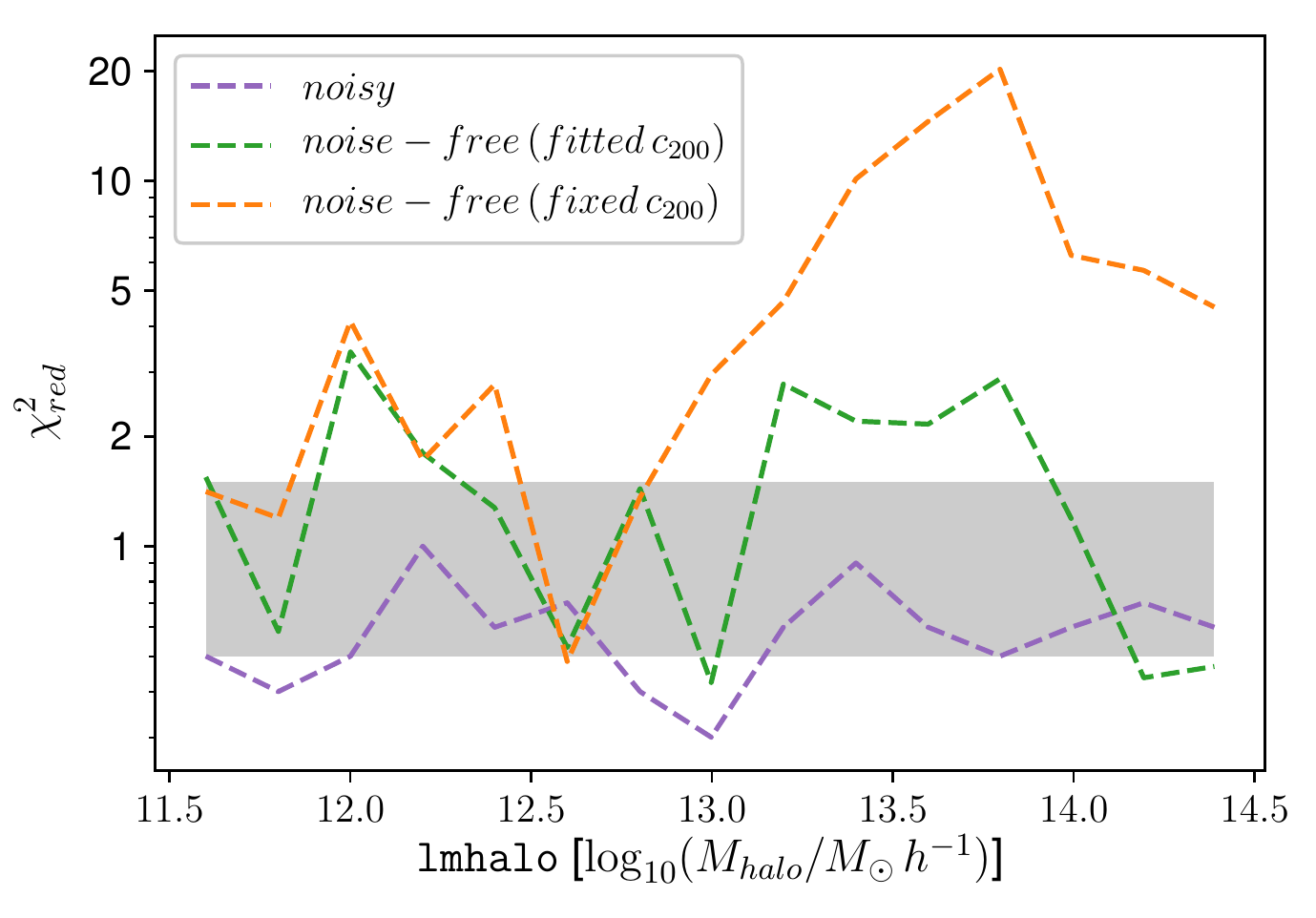}
    \caption{Reduced chi-square values, $\chi^2_{red}$, versus halo mass associated with the best-fitting NFW profiles computed considering the \textit{noise-free} source sample by fitting 
    the concentration parameter (dashed green line) and using the \citet{Duffy2008}
    relation to fit the profile (dashed orange line) and for profiles derived considering
    the \textit{noisy} source sample (dashed purple line). The gray region correspond to 
    $\chi^2$ values from 0.5 to 1.5.}
    \label{fig:chi2}
\end{figure}

According to the derived reduced chi-square values (Fig. \ref{fig:chi2}), 
lensing profiles are well constrained by a NFW model, except for masses \texttt{lmhalo} $> 13.5$\,log(M/M$_\odot$) computed by fixing the concentration parameters using the \citet{Duffy2008} relation.

\begin{figure}
    \centering
    \includegraphics[scale=0.65]{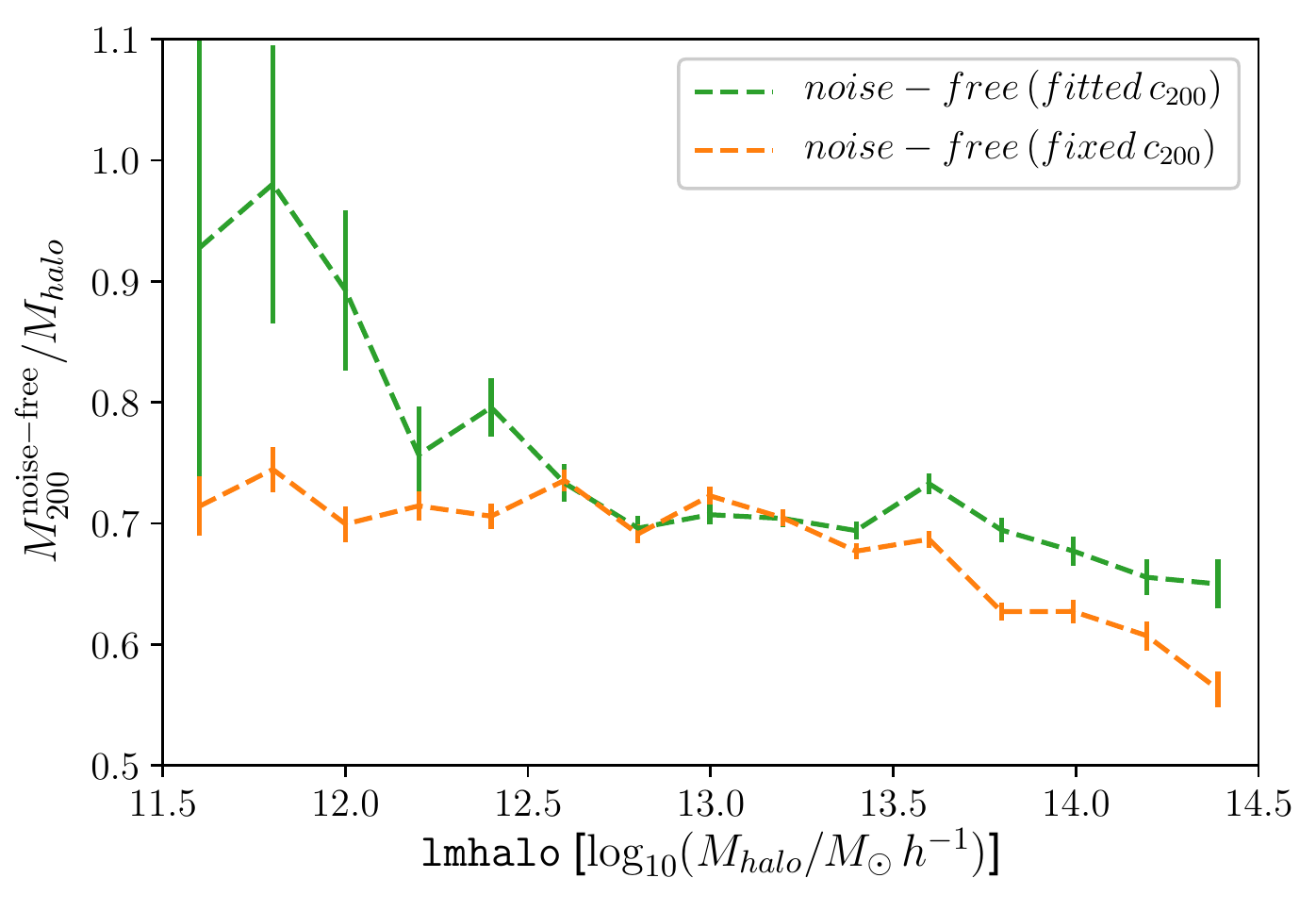}
    \caption{Ratio between the derived $M_{200}$ lensing masses and the 
    FOF halo masses derived from profiles obtained considering the \textit{noise-free}
    source sample, fitting the concentration parameter (dashed green line) 
    and considering the \citet{Duffy2008} relation to fit the profile (dashed orange line).}
    \label{fig:masses_8}
\end{figure}

Derived $M_{200}$ values based on the profiles using the \textit{noise-free}
sample source correlate well with the FOF halo masses,
\texttt{lmhalo}, see Fig.\ref{fig:masses_8}.
Note that we do not expect a one to one relation between FoF mass and $M_{200}$ \citep{White200,Jiang2014}.
In particular, we find that the estimated lensing mass to \texttt{lmhalo} ratio accurate follows a constant value $ \sim 0.7$ in the mass range $11.5-13.5 \, \log($M/M$ _\odot)$. For higher halo masses
(\texttt{lmhalo} $> 13.5$\,log(M/M$_\odot$)), lensing masses derived
by fitting the concentration parameter are larger than those obtained
when this parameter is fixed considering the \citet{Duffy2008} relation.
On the other hand, for lower halo masses (where the concentration 
parameter is poorly constrained) lensing masses have 
larger uncertainties and are lower than expected when the concentration is fixed.

\begin{figure}
    \centering
    \includegraphics[scale=0.65]{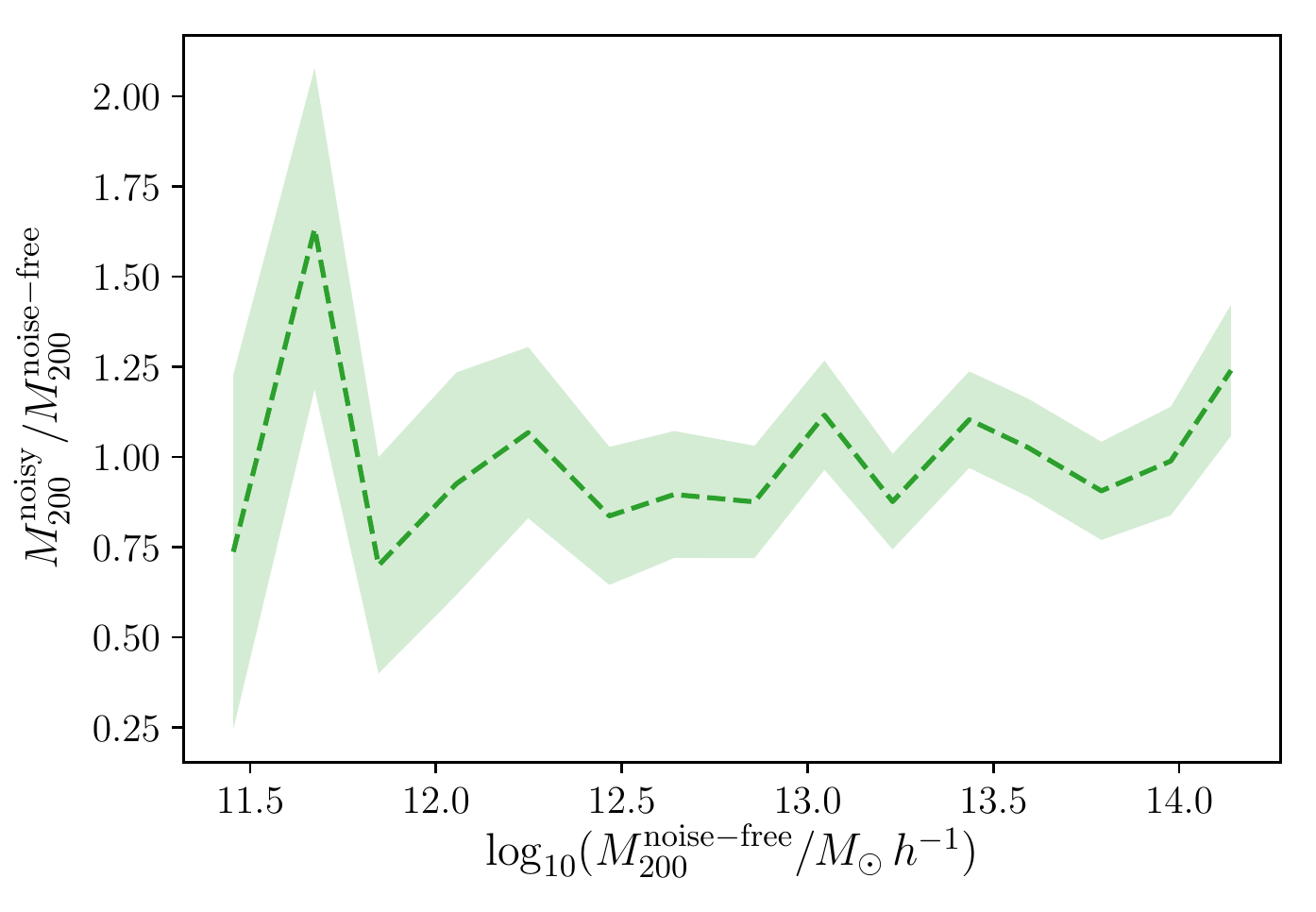}
    \caption{Ratio between derived lensing masses considering the \textit{noisy} and 
    the \textit{noise-free} source samples, taking into account the \citet{Duffy2008} relation.
    The shaded region corresponds to the fitted errors in the masses derived according to 
    the \textit{noisy} source sample.}
    \label{fig:masses}
\end{figure}

\begin{figure}
    \centering
    \includegraphics[scale=0.65]{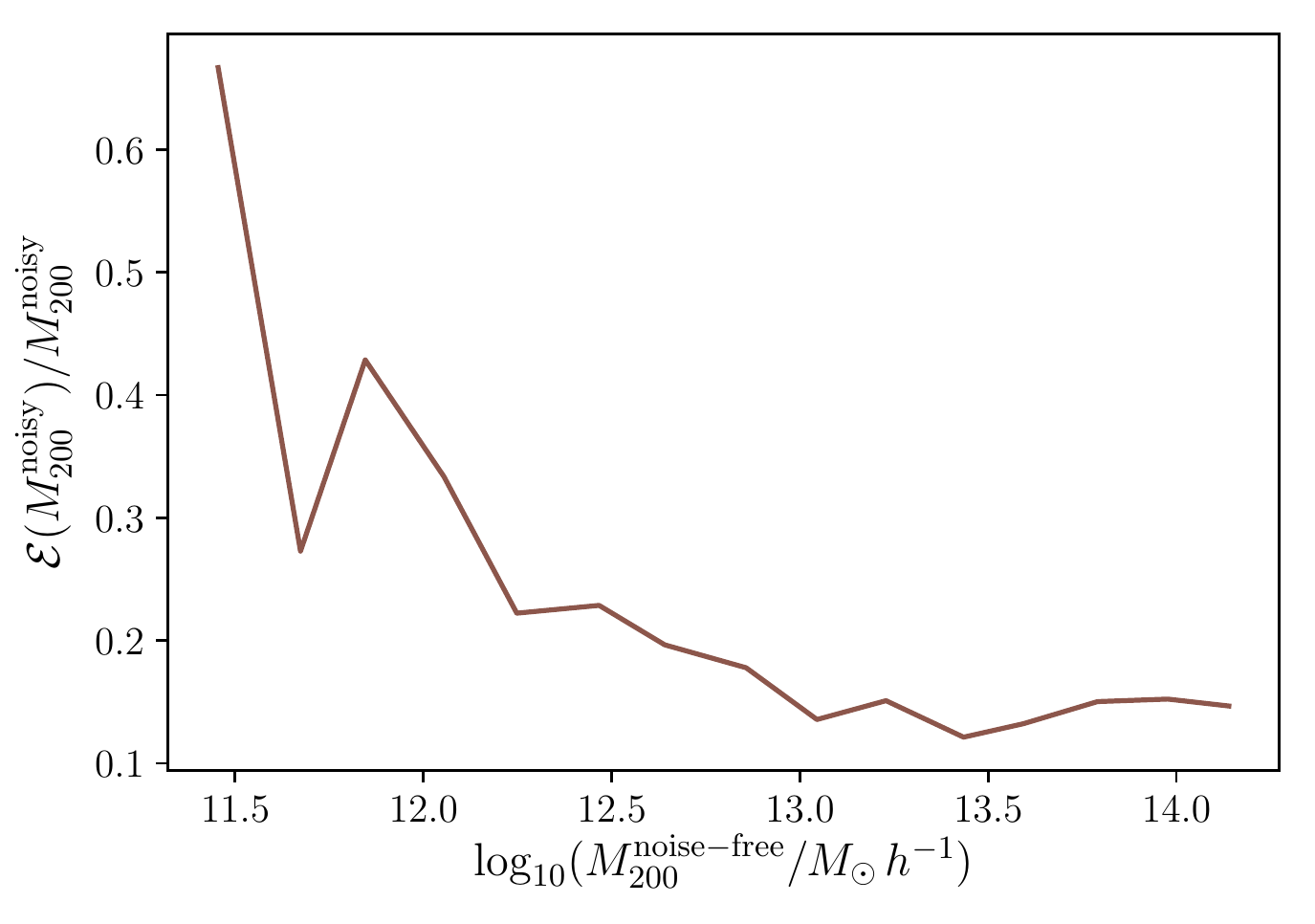}
    \caption{Lensing mass relative errors $\mathcal{E}(M_{200}^{noisy})/ M_{200}^{noisy} $ from the \textit{noisy} source sample versus the masses derived from the \textit{noise-free} source sample. }
    \label{fig:sn}
\end{figure}

As shown in Fig.\,\ref{fig:masses}, $M_{200}$ values derived for the \textit{noise-free}
source sample are consistent with those of the \textit{noisy} sample, $\langle M^{\mathrm{noisy}}_{200}/M^{\mathrm{noise-free}}_{200} \rangle = 1.00 \pm 0.22$. 
Nevertheless, the observed signal-to-noise ratio (S/N) for the \textit{noisy} source sample
drops significantly when considering halos with masses 
$< 12.5$\,log(M/M$_\odot$ h$^{-1}$). Figure \ref{fig:sn} shows
the relative mass uncertainty as a function of $M_{200}$ for the \textit{noise-free} sample. 
Halos with masses $> 12.5$\,log(M/M$_\odot$ h$^{-1}$)
can be detected with high significance. On the other hand,
inferred lensing masses for low
mass halos $\lesssim 12.0$\,log(M/M$_\odot$ h$^{-1}$) have a large uncertainty (relative error $> 30\%$).
This can also be seen by inspection 
of the density contrast profiles (Fig. \ref{fig:profiles}).
Lensing signal drops significantly down to the detection level for low-mass halos which
turns into underestimated masses. 
Taking into account the range of galaxy pair masses
(Fig. \ref{fig:massdist}) this effect can hamper the detection
of these galaxy systems. This issue is discussed in the next section. 

{}

{}

\subsection{Lensing analysis of galaxy pairs}

\begin{figure*}
    \centering
    \includegraphics[scale=0.65]{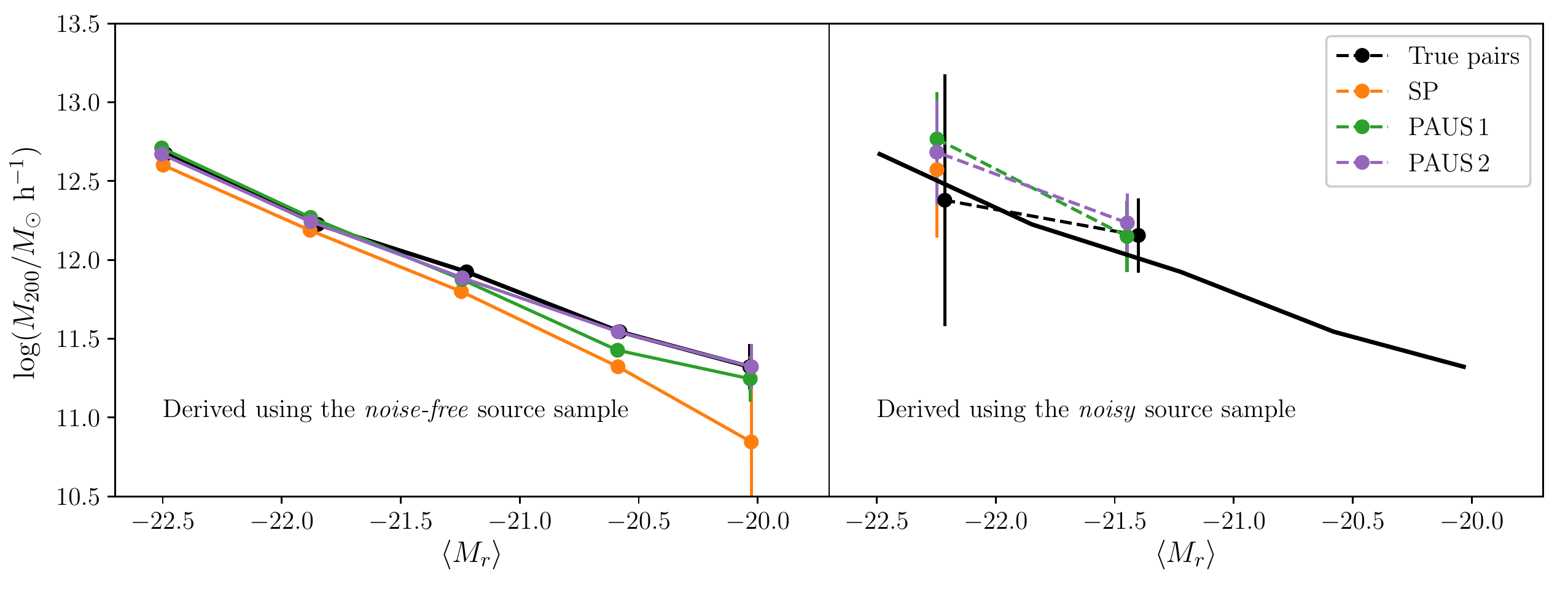}
    \caption{Derived lensing masses for the different pairs samples versus the
    average total magnitude of the pairs in the considered bin. In the left panel
    we show the result obtained by stacking the \textit{noise-free} source sample
    (solid lines). In the right panel we show the same relation in dashed lines
    for the masses obtained from the \textit{noisy} source sample and in solid 
    line for comparison the same relation as in the left panel obtained for the
    \textit{True pair} sample. For the higher magnitude bin in the case of the SP
    sample, the mass could not be properly constrained.}
    \label{fig:lummag_pairs}
\end{figure*}

We compute lensing masses for both the photometric pair
and the \textit{true pair} samples. By considering 
the \textit{noise-free} sources to compute the profiles,
we split the pairs according to their total $r-$band luminosity, $M_r$,
considering bins with an absolute magnitude width of 0.7.

The relation between the
derived lensing mass and mean luminosity in each bin is shown in Fig. \ref{fig:lummag_pairs}.
As can be seen, there is a good correlation between 
the estimated masses for the \textit{true pair} sample and the
\textit{photometric pairs}. This result reinforces the 
performance of our identification algorithm, which can 
properly recover the observational properties of galaxy pairs
located in isolated halos. However,
we notice that masses are systematically 
underestimated if larger errors in the photometric redshift
are considered for galaxy pairs with total absolute magnitudes
$M_r \gtrsim -21.5$. 

Reliable lensing masses from the \textit{noisy} source sample can only be obtained when considering galaxy pairs with high luminosity, $M_r \lesssim -21$. Taking into account the relation between the 
total absolute magnitude and halo mass \texttt{lmhalo} (Fig. \ref{fig:lummag_pairs}), this
threshold ensures that the masses are well constrained considering
the results presented in Fig.\,\ref{fig:sn}. 
Therefore only high luminosity pairs can be detected
with sufficient sensitivity taking into account the observational limitations.
For these high luminosity pairs we can recover the slope of the $M_{200}-M_r$
relation taking into account the photometric samples. 

For the highest luminosity bin of the \textit{true pair} sample, we obtain a larger uncertainty due to the low number of pairs in this bin.
Thus, using a low number of sources, the stacking procedure lacks effectiveness in providing suitable mass estimates. We notice, however, that the general trend of the mass-luminosity relation is well recovered.

\begin{figure*}
    \centering
    \includegraphics[scale=0.7]{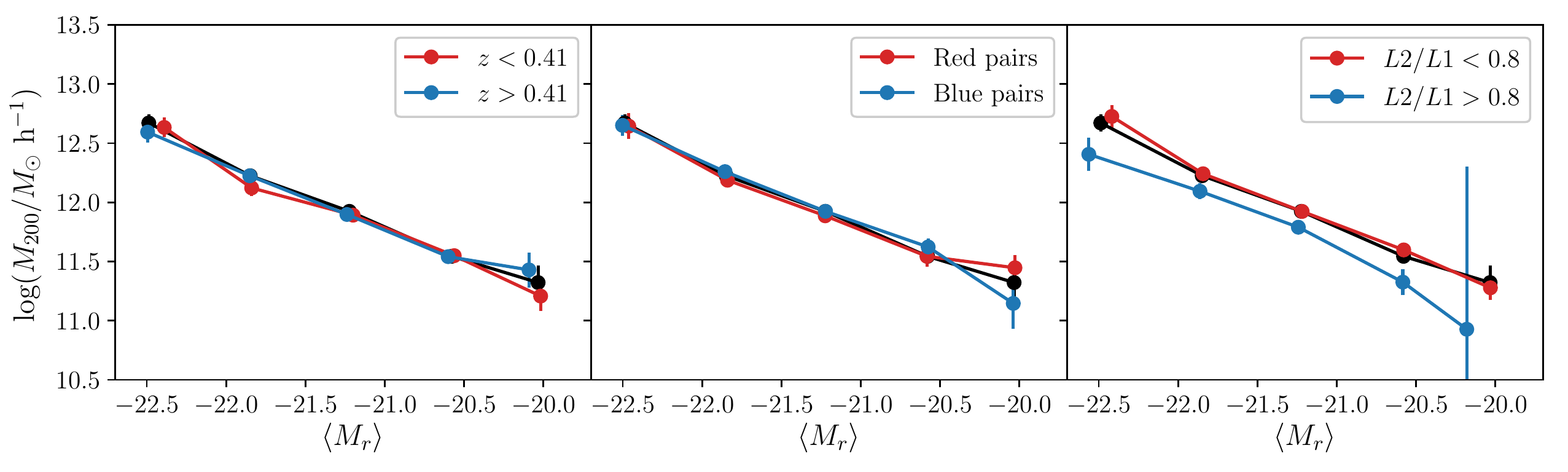}
    \caption{Derived lensing masses for all the \textit{True pairs} versus the
    average total magnitude of the pairs in the considered bin obtained
    according to the \textit{noise-free} source sample. In red and blue we show 
    the same relation but splitting the \textit{True pairs} according to 
    the pair redshift (left panel), color (middle panel) and luminosity ratio
    of the galaxy members.}
    \label{fig:lummag_pairs_2}
\end{figure*}

We have also explored the dependence of the mass-luminosity relation by adopting different selection criteria for the pair samples taking into account redshift, color and luminosity ratio of member galaxies, see Fig, \ref{fig:lummag_pairs_2}. We find no difference in the mass-luminosity relation for samples selected according to the median redshift of the sample pairs ($z = 0.41$) as seen in the left panel of Fig. \ref{fig:lummag_pairs_2}. We perform the red/blue galaxy classification taking into account the redshift and absolute magnitude bins of the member galaxies, finding no significant differences
between the red and blue populations (see Fig.\ref{fig:lummag_pairs_2} middle panel). This result contrasts with the finding of \cite{Gonzalez2019}, where they obtain red pairs exhibiting larger lensing masses. Further analysis needs to be performed in order to determine whether these discrepancies can be explained from colour-density dependence difference between the simulation and the observations. Galaxy luminosities and colours are assigned in the MICE mock catalog in order to match observed galaxy properties and the clustering dependence on these parameters \citep{Carretero2015}. In particular galaxy colors are assigned to fit the observed ($g -r$) vs. $M_r$ SDSS observed relation \citep{Blanton2003} and the clustering properties as a function of color \citep{Zehavi2011}. The procedure is similar to the model presented in \citet{Skibba2009} in which colors depend on galaxy type (whether it is a central or a satellite galaxy), and on its color sequence (red, blue and green), but not on the parent halo mass. Firstly, colors are assigned to the satellite galaxies considering their absolute magnitudes to set the fraction of satellite galaxies that belongs to the red
and green sequences. Since the systems analyzed in this work reside in low-mass halos, the number of expected satellites is small and so a high uncertainty is expected in its color assignment leading to possible discrepancies with observed colors for these particular systems.

We note, however, that the results of \cite{Gonzalez2019} results may be explained by the inclusion
of unbound systems in the blue sample of pairs, resulting in  lower derived  lensing masses. In the right panel of Fig.\ref{fig:lummag_pairs_2}, we explore the dependence of the total mass-luminosity relation of the pairs selected according to the pair member luminosity ratio, $L2/L1$. It can be seen that the estimated lensing masses are about two times smaller for pairs with similar luminosity members. Although this result can not be tested with the present observational data, since
the masses are poorly estimated when considering noise, this result
could be addressed with better quality data in future surveys. 

\section{Summary and conclusions}
\label{Sum}

We present an algorithm for the identification of galaxy pairs based on photometric information. The identification algorithm successfully reproduces the distribution of total luminosity and mass of truly bound galaxy pairs residing in the same dark matter halo. Pairs are identified through a commonly used procedure: adopted fixed values of projected separation $r_p$ and relative velocities $\Delta V$ plus the requirement that all systems have at least one bright member ($M_r < -19.5$) and that the pair members have an apparent magnitude difference $ \Delta m<2$. Finally, we apply an isolation criterion that allows us to exclude pairs in massive systems. The algorithm was applied to three galaxy samples from the MICE simulation that consider different photometric redshifts uncertainties, and derive three catalogs of \textit{photometric pairs} in the redshift range $0.2-0.6$. 

In order to test our identification algorithm, we select galaxy pairs that meet all the criteria described above and additionally,  both galaxy members reside in the same halo, labeled as  the \textit{true pairs} sample. This restriction provides a novel approach to testing galaxy pair identification techniques. Then, we compare the recovered photometric pairs with the \textit{true pairs}. As expected, we find that the identification reliability improves as the photometric redshift error decreases. Nevertheless, all the pair samples identified based on photometric data recover properly the distribution of observational properties of the \textit{true pairs}, namely total luminosity and pairs members luminosity ratio. The derived luminosity ratio of the photometric pair members shows a bi-modal behavior. Also, we find that pairs with similar luminosity members (larger luminosity ratios) tend to reside in more massive halos at a given total pair luminosity. As the accuracy in the photometric redshift error improves, galaxy pair samples tend to recover more systems classified as  \textit{true pairs}. This ensures that PAUS will provide a valuable contribution for the identification of galaxy systems. 

We have also studied the different  pair samples using weak lensing techniques. In order to test the performance of our lensing analysis and its ability to recover total halo masses, we first analysed a sample of halos within the same redshift range as the pairs, and with halo masses larger than $> 11.5$\,log(M/M$_\odot$). Source galaxies were selected considering the CFHTLenS data properties in order to mimic observational conditions. Lensing masses were obtained applying stacking techniques by splitting the total halo sample into different halo mass bins. We find that derived density contrast profiles of higher mass halos are less concentrated than predicted by \citet{Duffy2008}. For lower mass halos, the lensing analysis cannot properly constrain the halo concentration parameter. Nevertheless, the derived $M_{200}$ strongly correlates with the total halo FOF masses provided by MICE. When considering source samples with lensing properties adding observational noise, the concentration parameter cannot be accurately determined, but  the derived $M_{200}$ is in excellent agreement with the values obtained without observational noise. However, the signal-to-noise significantly drops when considering halos with masses $< 12$\,log(M/M$_\odot$).

 Although lensing masses tend to be systematically underestimated for the samples with larger photo-z errors, in general, masses for all the identified samples are successfully recovered with our analysis. Even when observational noise is considered for the lensing analysis, we can successfully recover the slope of the lensing mass versus total luminosity relation. However, masses can be determined only for galaxy pairs with total absolute $r-$band magnitudes brigther than $-21$. This luminosity threshold roughly corresponds to galaxy pairs in halos with masses $> 12.5$\,log(M/M$_\odot$). When considering galaxy pairs identified using standard photometric redshift uncertainties (i.e. a factor 2.7 higher than the typical error predicted for PAUS) the lensing signal is lowered, since less galaxy pairs are identified. For this photometric selected sample, lensing masses can only be recovered for the most luminous pairs. It is important to highlight that although the selection criteria for the galaxy pair identification can be relaxed, by considering larger limits for $\Delta V$ or $\Delta m$, this would result in a decrease in the purity of the selected sample albeit with no improvement of the lensing signal.
 
The results obtained show that the upcoming PAUS data with high quality photometric redshift information will enable the construction of large and reliable samples of galaxy systems. Isolated galaxy pair identification is a challenging task since these low mass systems can only be detected based on a low number of photometric parameters. In this sense, it is important to apply accurate tests that ensure the recovery of truly bound systems. The present algorithm allows us to obtain suitable samples that can be used to obtain physical properties leading to a deeper understanding of their formation and evolution in a cosmological context.

\begin{acknowledgements}

We thank to the anonymous referee for his/her comments that helped to improve this work. 

We thank Carlton Baugh for his careful reading of the manuscript that has helped us to improve the text.

This project has received funding from the European Union’s Horizon 2020 Research and Innovation Programme under the Marie Sklodowska-Curie grant agreement No 734374.

This work has been supported by MINECO  grants AYA2015-71825 \& ESP2015-66861. IEEC is partially funded by the CERCA program of the Generalitat de Catalunya.

This work was also partially supported by Agencia Nacional de Promoci\'on Cient\'ifica y Tecno\'oogica (PICT 2015-3098), the Consejo Nacional de Investigaciones Cient\'{\i}ficas y T\'ecnicas (CONICET, Argentina) and the Secretar\'{\i}a de Ciencia y Tecnolog\'{\i}a de la Universidad Nacional de C\'ordoba (SeCyT-UNC, Argentina).

MS has been supported by the European Union's  Horizon 2020 research and innovation programme under the Maria Skłodowska-Curie grant agreement No 754510 and National Science Centre (grant UMO-2016/23/N/ST9/02963).

MM acknowledges support from the Beatriu de Pinos fellowship (2017-BP-00114).

We made an extensive use of the following python libraries:  http://www.numpy.org/, http://www.scipy.org/, https://www.astropy.org/ and http://www.matplotlib.org/.

This work has made use of CosmoHub.
CosmoHub has been developed by the Port d'Informació Científica (PIC), maintained through a collaboration of the Institut de Física d'Altes Energies (IFAE) and the Centro de Investigaciones Energéticas, Medioambientales y Tecnológicas (CIEMAT), and was partially funded by the "Plan Estatal de Investigación Científica y Técnica y de Innovación" program of the Spanish government.

EJG and FR would like to specially thank the contribution of Dar\'io Gra\~{n}a who helped to run the algorithms. Gracias por las picadas. Go team Barcelona!

EJG, FR and AO thank Orion for the laughter and the good luck.

\end{acknowledgements}


\bibliography{references}

\end{document}